\documentclass[aps,prl,twocolumn,10pt,showpacs,superscriptaddress,amsmath,amssymb]{revtex4-1}

\usepackage{graphicx} 
\usepackage{dcolumn} 
\usepackage{bm} 
\usepackage{nicefrac} 
\usepackage{xspace} 

\DeclareMathOperator{\sech}{sech} 
\DeclareMathOperator{\csch}{csch} 

\newcommand{\NLS}{NLS\xspace}
\newcommand{\GP}{GP\xspace}
\newcommand{\BdG}{BdG\xspace}
\newcommand{\bigSoliton}{original soliton\xspace}

\def\px{\partial_x} 

\begin{document}

\title{Emergence of Reflectionless Scattering from Linearizations of Integrable PDEs \\ around Solitons}

\author{Andrew Koller}
\affiliation{Department of Physics, University of Colorado, Boulder, CO 80309, USA}
\author{Zaijong Hwang}
\affiliation{Department of Physics, University of Massachusetts Boston, Boston, MA 02125, USA}
\author{Maxim Olshanii} \email{Maxim.Olchanyi@umb.edu}
\affiliation{Department of Physics, University of Massachusetts Boston, Boston, MA 02125, USA}

\date{November 15, 2014}


\begin{abstract}
We present four examples of integrable partial differential equations (PDEs) of mathematical physics that---when linearized around a stationary soliton---exhibit scattering without reflection at {\it all} energies. Starting from the most well-known and the most empirically relevant phenomenon of the transparency of one-dimensional bright bosonic solitons to Bogoliubov excitations, we proceed to the sine-Gordon, Korteweg-de Vries, and Liouville's equation whose stationary solitons also support our assertion. The proposed connection between integrability and reflectionless scattering seems to span at least two distinct paradigms of integrability: S-integrability in the first three cases, and C-integrability in the last one. We argue that the transparency of linearized integrable PDEs is necessary to ensure that they can support the transparency of stationary solitons in the original integrable PDEs. As contrasting cases, the analysis is further extended to cover two non-integrable systems: a sawtooth-Gordon and a $\phi^4$ model.
\end{abstract}

\maketitle

\section{Introduction} It is from the field of fiber optics that we know that small excitations of a soliton of the nonlinear Schr\"{o}dinger (NLS) equation---also called the Gross-Pitaevskii (GP) equation in other contexts---can penetrate the \bigSoliton without any reflection \cite{kaup1990_5689}. An analogue of the effect, the transparency of the bosonic Bogoliubov-de Gennes (\BdG) equation, was later identified \cite{castin2009_317} for solitons in Bose condensates \cite{khaykovich2002_1290,strecker2002_150} as well.

Potentials that are transparent at all energies have been studied extensively since the 1933 discovery of the P\"oschl-Teller potential \cite{flugge1994}
	\begin{align}
	V(x) = -\tfrac{1}{2} n (n+1) \sech^2(x)\,\label{PT}
	\end{align}
with integer $n$ for the time-independent Schr\"{o}dinger equation. Several other classes of physically relevant linear differential equations have also been found to support reflectionless potentials  \cite{sukumar1985_2917,sukumar1986_2297,barclay1993_2786,cooper1988_1,cooper1988_145} (see also book \cite{susy_book} for more references).

Reflectionless potentials also appear in the context of finding exact solutions to integrable nonlinear PDEs with the inverse scattering transform. The process of obtaining exact solutions to such S-integrable systems is greatly simplified if the corresponding Lax Liouvillian is transparent. Regardless of transparency, the bound states of Lax Liouvillians will also manifest as solitons in the original nonlinear PDE \cite{solitons_book}. The reflectionless property of the solitonic \BdG Liouvillian---a linearized NLS equation---is not related to a Lax Liouvillian however. This implies that an independent connection between integrability and reflectionless scattering may be at work.

We will begin our study by examining the \BdG equation in detail, from which we may formulate a conjecture on the nature of this independent connection between integrability and reflectionless scattering. Three more examples will follow---involving the sine-Gordon (sG), Korteweg-de Vries (KdV), and Liouville's equations---that confirm this connection. For a contrasting study, the same analysis will also be performed on two non-integrable nonlinear PDEs: a sawtooth-Gordon equation and a $\phi^4$ model.


\begin{figure*} 
\centering
\begin{tabular}{cc}
\includegraphics[scale=.5]{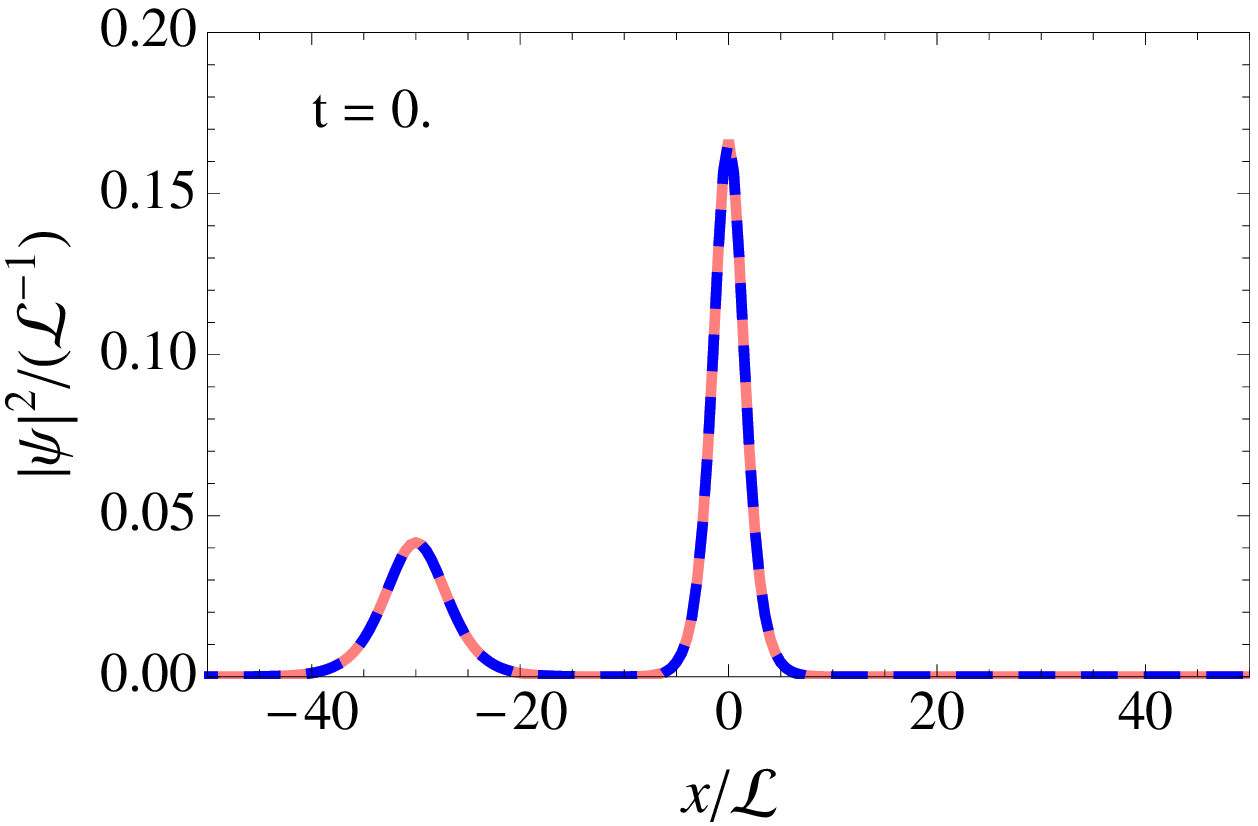} & \includegraphics[scale=.5]{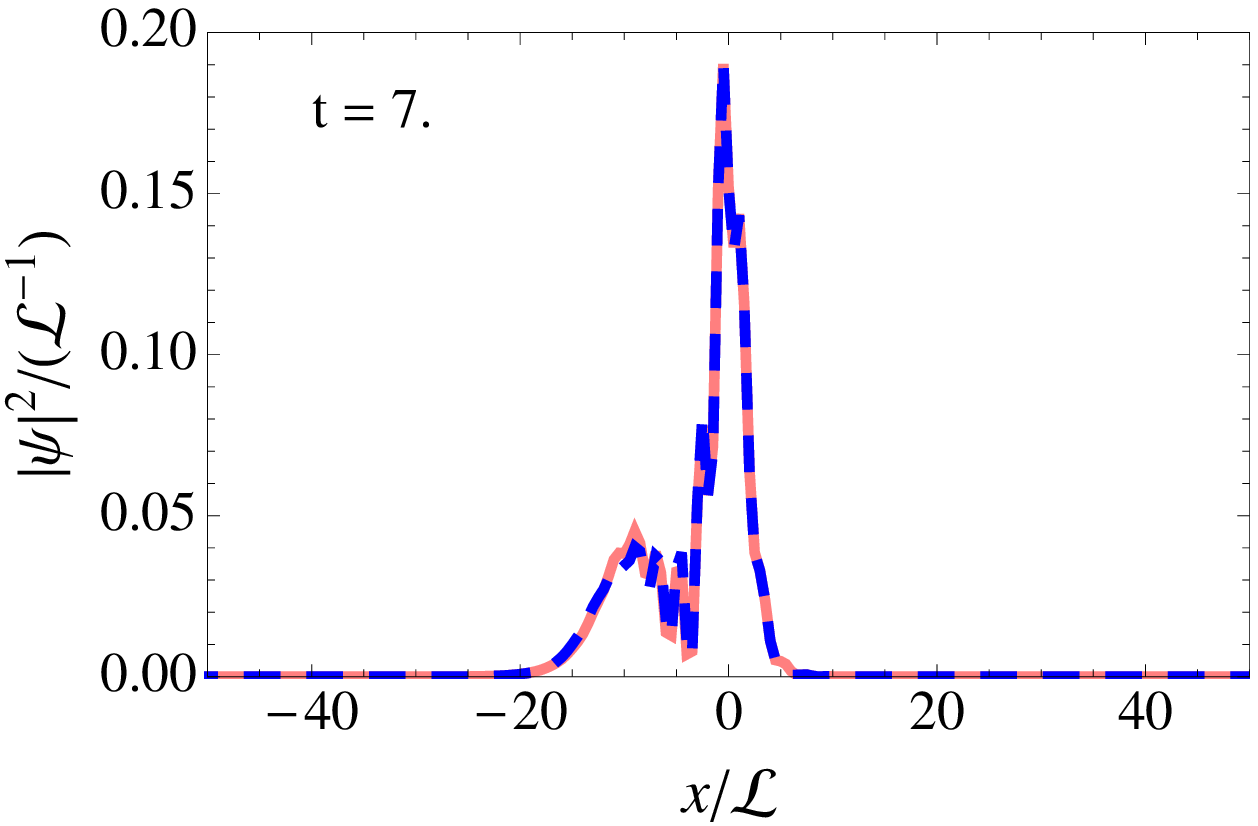}\\
\includegraphics[scale=.5]{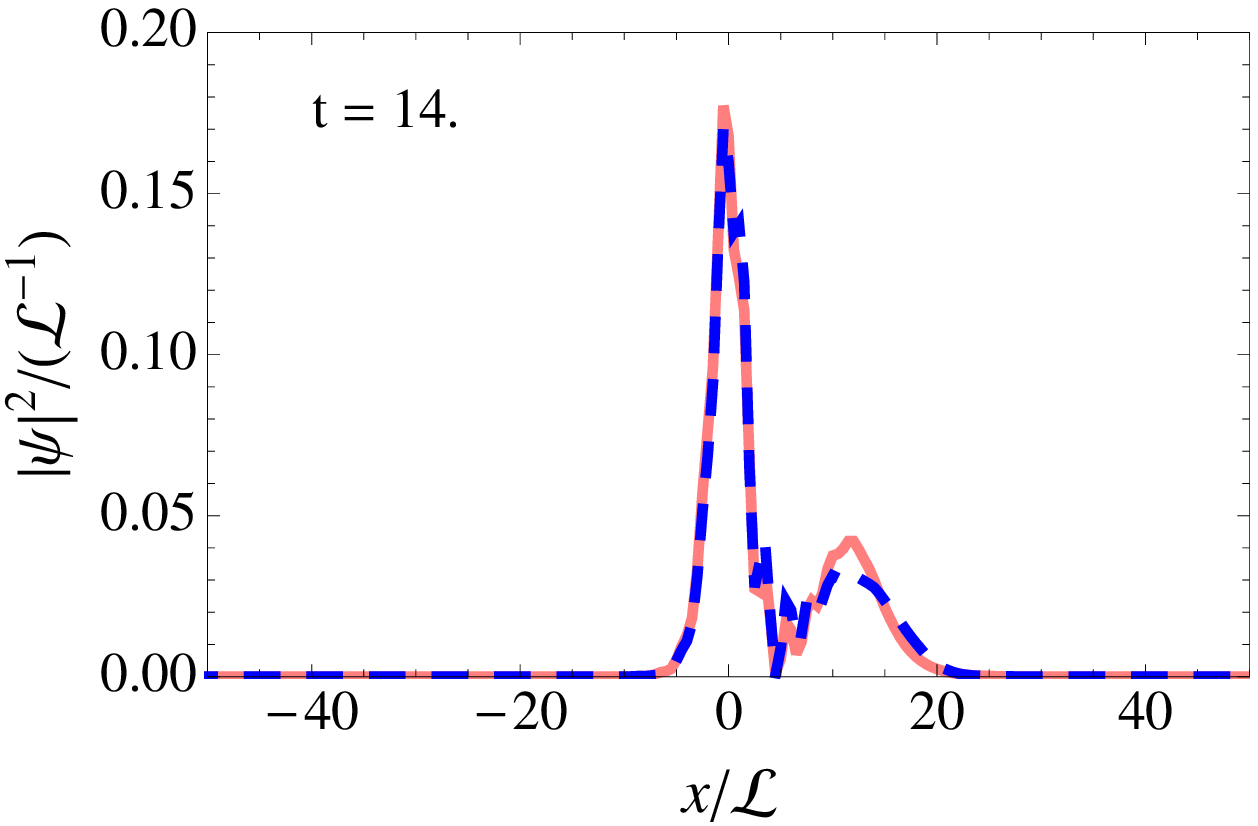} & \includegraphics[scale=.5]{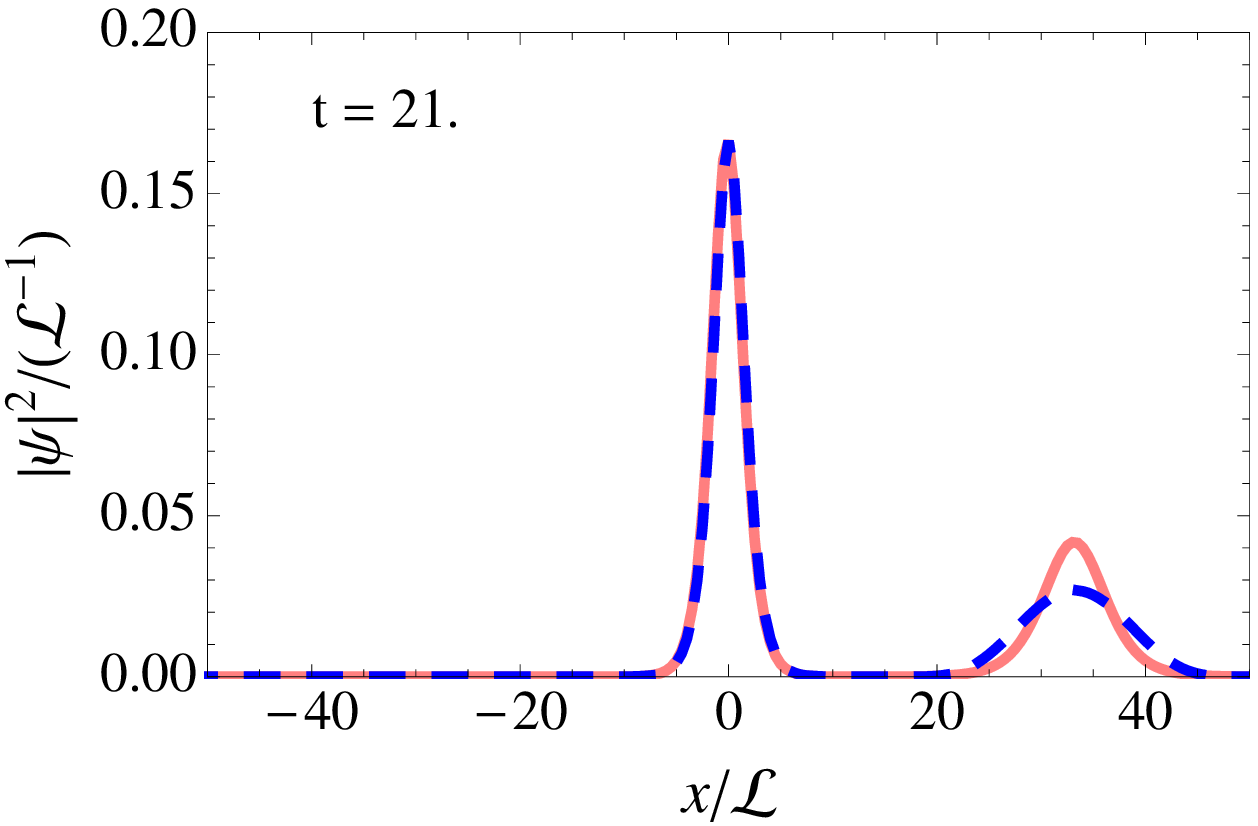}
\end{tabular}
\caption
{
\label{fig:two_NLS_and_BdG_solitons}
A simulation illustrates how the Bogoliubov-de Gennes (BdG) equation, which is a linearized version of the one-dimensional attractive Gross-Pitaevskii (GP) equation (i.e. a 1D attractive nonlinear Schr\"odinger (NLS) equation), is capable of describing the penetration of a smaller soliton through a larger one without reflection. This observation indicates that the BdG equation itself must show the property of scattering without reflection. In this example, the population of the smaller soliton is half that of the larger one. Although this hardly qualifies as a {{\it small}} perturbation, it helps to accentuate the difference between the exact solution and its linearized counterpart, while also demonstrating the robustness of the result with respect to the size of the solitons involved.
Solid red line: the solution of the \GP equation. Dashed blue line: the solution of the \BdG equation.
The units of length and time are ${\cal L} \equiv \hbar^2/(m g N) = \ell/2$ and ${\cal T} \equiv \hbar^3/(m g^2 N^2) = \hbar/(8|\mu_{0}|)$, respectively.
}
\end{figure*}


\section{Linearization of\\attractive \NLS equation} Consider an attractive \NLS equation (also called the GP equation in the context below) describing the mean-field dynamics of an ensemble of $N$ one-dimensional bosons with attractive $\delta $-interactions
	\begin{align}
	i\hbar\psi_t = \left( -\tfrac{\hbar^2}{2m}\px^2 - gN\left|\psi\right|^2 \right) \psi\,,
	\label{GP}
	\end{align}
where $g$ is the coupling constant and $m$ is the particle mass. $\sqrt{N} \psi(x,\,t)$ constitutes a mean-field approximation for the bosonic quantum field $\hat{\Psi}(x,\,t)$, and the normalization condition is $\int_{-\infty}^{+\infty}\! dx \, |\psi(x,\,t)|^2 = 1$ accordingly. Equation (\ref{GP}) is known to support solitons \cite{solitons_book} that have been successfully observed in experiments \cite{khaykovich2002_1290, strecker2002_150}.

An example of a two-soliton collision between a small moving soliton and a large stationary one is shown in Fig.\nobreak\ \ref{fig:two_NLS_and_BdG_solitons}. Assuming that the moving soliton is sufficiently small, this collision can be approximately described by a \BdG linearization around the larger soliton as
	$$
	\psi(x,\,t) \approx \bar{\psi}(x,\,t) + \delta\psi(x,\,t)\,.
	$$
Let the larger soliton be
	$$
	\bar{\psi}(x,\,t) = \bar{\phi}(x) \exp(-i\mu_{0}t/\hbar)\,,
	$$
that originates from a stationary solitonic solution
	$$
	\bar{\phi}(x) = (2 {\ell})^{-\nicefrac{1}{2}} \sech(x/\ell)\,,
	$$
of the time-independent \NLS equation
	$$
	\bigl( -\tfrac{\hbar^2}{2m}\px^2 - gN|\bar{\phi}|^2 \bigr) \bar{\phi} = \mu_{0} \bar{\phi}\,.
	$$
In the equations above,
	$
	\ell = 2 \hbar^2/(m g N)
	$
is the size of the soliton, and
	$
	\mu_{0} = - m(gN)^2/(8\hbar^2)
	$
is its chemical potential. Direct substitution into equation (\ref{GP}) then produces to first order the \BdG equation
	\begin{align}
	&i\hbar\,\delta\psi_t = \left( -\tfrac{\hbar^2}{2m}\px^2 - 2 gN|\bar{\psi}|^2 \right) \delta\psi - gN\bar{\psi}^2\,\delta\psi^*\,. \label{BdG}
	\end{align}
As demonstrated by the simulation shown in Fig.\nobreak\ \ref{fig:two_NLS_and_BdG_solitons}, the \BdG equation above properly supports the transparency of the larger soliton to the smaller one. This implies that the stationary scattering solutions of the equation must exhibit transparency as well.

Let us now verify the simulation in Fig.\nobreak\ \ref{fig:two_NLS_and_BdG_solitons} by displaying the scattering solutions to the BdG equation (\ref{BdG}) explicitly. To simplify the formulas that follow, we will use a system of units with
	$
	\hbar = m = \ell = 1
	$.
The \NLS equation thus becomes
	$$
	i\,\psi_t = \bigl( -\tfrac{1}{2}\px^2 - 2\left|\psi\right|^2 \bigr) \psi\,,
	$$
and the solitonic solution is
	$$
	\bar{\psi}(x,\,t) = \left(\nicefrac{1}{\sqrt{2}}\right) \sech x\cdot\exp(i\,t/2)\,,\\
	$$
with
	$
	\mu_{0} = - \nicefrac{1}{2}
	$.
Using the same linearization of
	$$
	\psi(x,\,t) \approx \bar{\psi}(x,\,t) + \delta\psi(x,\,t)\,,
	$$
we find the time dependent \BdG equation
	$$
	i\,\delta\psi_t = \bigl( -\tfrac{1}{2}\px^2 - 4 |\bar{\psi}|^2 \bigr)\,\delta\psi - 2 \bar{\psi}^2\,\delta\psi^*\,,
  $$
that has scattering solutions of
	$$
  \delta\psi(x,\,t) = \left(u_{k}(x) e^{-i\epsilon t} + (v_{k}(x))^* e^{+i\epsilon t} \right) e^{-i\mu_{0}t}\,,
  $$
where the two-component wavefunctions $(u_{k}(x),\,v_{k}(x))^\intercal$ are the positive energy eigenstates of a \BdG Liouvillian,
	\begin{align}
		\hat{H} =
		\begin{pmatrix}
		-\frac{1}{2}\px^2 - 2\sech^2\!x + \frac{1}{2} & -\sech^2\!x \\
		\sech^2\!x & \frac{1}{2}\px^2 + 2\sech^2\!x - \frac{1}{2}
		\end{pmatrix} \,.
		\label{BdG_Liouvillian}
	\end{align}
These eigenstates have the form \cite{kaup1990_5689}
	\begin{align}
		\begin{pmatrix}
			u_{k}(x) \\ 
			v_{k}(x)
		\end{pmatrix}
		\propto
    \begin{pmatrix}
    	(1+i k^{-1}\tanh^2\!x)^2 \\
    	k^{-2} \sech^2\!x
		\end{pmatrix}
		\exp(i k x) \,,
		\label{BdG_u_and_v}
	\end{align}
and satisfy the eigenvalue equation
	$$
  \hat{H} \begin{pmatrix} u_{k} \\ v_{k} \end{pmatrix}
  =
  \epsilon(k) \begin{pmatrix} u_{k} \\ v_{k} \end{pmatrix},
  $$
  $$
  \epsilon(k) = \tfrac{1}{2}k^2 -\mu_{0} = \tfrac{1}{2}k^2 + \tfrac{1}{2},
  $$
and $-\infty < k < +\infty$ \cite{full_BdG}. As expected, the scattering solutions in equation (\ref{BdG_u_and_v}) show no reflection off the soliton, at {\it all} energies.

A direct calculation of the scattering solutions is not the only way to verify that the BdG equation (\ref{BdG}) supports scattering without reflection however. An elegant algebraic explanation is also provided by connecting $\hat{H}$ to a scatterer-free Liouvillian
	\begin{align}
	\hat{H}_{0} &=
		\left( \begin{array}{cc}
		-\frac{1}{2}\px^2 + \frac{1}{2} & 0 \\
		0 & \frac{1}{2}\px^2 - \frac{1}{2}
		\end{array} \right),
	\label{BdG_H0}
	\end{align}
through an intertwining relationship
	\begin{align}
	\hat{H} \hat{\cal A} = \hat{\cal{A}} \hat{H_0}
	\label{intertwining_relationship}
	\end{align}
where the intertwiner $\hat{\cal{A}}$ is represented by \cite{non_uniquiness}
	\begin{align}
	\hat{\cal A} &=
		\left( \begin{array}{cc}
		\hat{F} & \hat{G} \\
		\hat{G} & \hat{F}
		\end{array} \right),
		\label{BdG_intertwiner}	\\
	\hat{F} &= \px^4 + (1-2\tanh x)\,\px^3 + (\tanh x-1)^2\,\px^2 \nonumber \\
		&\hphantom{={}} + (1 - \sech^2\!x-2\tanh x)\,\px + \tanh^2\!x\,,\nonumber \\
	\hat{G} &= -\sech^2\!x\,(\px^2 + \px + 1)\,.\nonumber
	\end{align}
Now we can observe that the eigenstates $w_{k}(x) \equiv (u_{k}(x),\,v_{k}(x))^\intercal$ of $\hat{H}$ may be obtained from the eigenstates $w_{k,\,0}(x)$ of $\hat{H_0}$ through the map
	$$
	w_{k}(x) \propto \hat{\cal{A}}\,w_{k,\,0}(x).
	$$
The spectra of $\hat{H}$ and $\hat{H_0}$ are therefore identical up to the kernel of $\hat{\cal{A}}$.

Since $\hat{H}_{0}$ is a differential operator with constant coefficients, and $\hat{\cal{A}}$ is a differential operator with coefficients that tend to a constant value as $x \to \pm\infty$, the plane-wave eigenstates
	$
	w_{k,\,0}(x) = (1,\,0)^\intercal\exp(i k x)
	$
of $\hat{H_0}$ \cite{comment_on_antiparticles} will give rise to a single-plane-wave asymptotic behavior of the eigenstates of $\hat{H}$, such that
	$
	w_{k}(x) \propto \exp(i k x)
	$
as $x \to \pm\infty$ too. This result is precisely scattering without reflection, and it is fully consistent with the explicit scattering solutions in (\ref{BdG_u_and_v}).

The intertwining relationship (\ref{intertwining_relationship}) and its associated supersymmetric (SUSY) algebra have been studied extensively; a comprehensive review can be found in \cite{andrianov2012_503001}. The presence of an intertwiner connecting a reflectionless Liouvillian $\hat{H}$ to its asymptotically constant and translationally invariant form $\hat{H_0}$ is a generic mathematical mechanism that underlies reflectionless scattering at all energies \cite{sukumar1986_2297}. Indeed, for all four examples of integral nonlinear PDEs presented in this paper, it is possible to start from a reflectionless Liouvillian candidate $\hat{H}$, find its asymptotic form $\hat{H_0}$, solve the intertwining relation (\ref{intertwining_relationship}) directly for the intertwiner, then construct the scattering solutions of $\hat{H}$ by applying the intertwiner to the eigenstates of $\hat{H_0}$.


\section{Emergence of\\reflectionless scattering}

Through numerical simulation, explicit scattering solutions, and an intertwining relationship, the preceding section has confirmed that a linearization of the NLS equation---the BdG equation---admits scattering without reflection, but it has not resolved the question of {\it why} the \BdG equation should be reflectionless. Given the rarity of both integrable systems and reflectionless potentials, it would be natural to suspect a connection between the two, yet the reflectionless property of the \BdG equation is not of the familiar type associated with a Lax Liouvillian. The \BdG equation is merely the linearization of the NLS equation, and not the associated scattering problem that one would construct for the NLS under an inverse scattering transform.

To understand the connection between integrability and reflectionless scattering for the \BdG equation, we must first differentiate solitons from solitary waves in the current context. While solitons are also solitary waves---both can propagate without dispersion---solitons possess the additional ability to penetrate each other without loss of identity. This unique property enshrines solitons as a hallmark of integrable systems and distinguishes them from solitary waves that can also exist in non-integrable systems.

The ability of small solitons to penetrate large ones without reflection is thus a direct consequence of the integrability of the system. From physical intuition, we may reasonably expect this transparency to be preserved even if the underlying integrable PDE were linearized, provided that the small solitons are {\it small enough} in the appropriate sense. If a large soliton is required to be transparent to a small soliton of {\it any} size and velocity in the linearized equation, and the small soliton can be represented as a wave packet that is a superposition of scattering solutions, then we can conjecture that the large soliton should also be transparent to incident plane waves of {\it any} wavelength. In other words, the linearization of an integrable PDE around a soliton should produce scattering without reflection at all energies. Let us test this conjecture now with three integrable equations.


\section{Linearization of\\sine-Gordon equation} Linearization of the sine-Gordon equation
	\begin{align*}
  	u_{xx} - u_{tt} = \sin u
  	\end{align*}
may at first appear ill-defined since the size of the solitons here is quantized, so they can never be small. Nonetheless, the linearized sine-Gordon equation must still be transparent in order to support the ability of small moving breathers to penetrate a stationary kink-soliton in the form of \cite{solitons_book}
	\begin{align*}
	\bar{u}(x) = 4 \arctan e^x\,.
	\end{align*}
Consider now an approximate solution that consists of the soliton and a small, time-periodic perturbation,
	$$
	u(x,\,t) \approx \bar{u}(x) + \delta u(x)\,e^{-i\omega t}\,.
	$$
To the first order of perturbation theory, the excitation $\delta u(x)$ will obey
	\begin{align}
	\bigl( -\px^2 - 2\sech^2\!x \bigr)\, \delta u  = E \, \delta u ,
	\label{PT_equation}
	\end{align}
with $E=\omega^2-1 $, yet this is nothing else but the celebrated P\"oschl-Teller equation (\ref{PT}) with $n=1$ whose scattering solutions correspond to {\it pure transmission without reflection} \cite{flugge1994}. That the P\"oschl-Teller Hamiltonian,
	$
	\hat{H} \equiv -\px^2 - 2\sech^2\!x
	$,
is reflectionless can easily be verified by intertwining it, in the same manner as equation (\ref{intertwining_relationship}), 
with its scatterer-free counterpart
	$
	\hat{H_0} = -\px^2
	$
using the intertwiner
	$
	\hat{\cal{A}} = -\px+\tanh x 
	$ 
\cite{sukumar1985_2917}. The corresponding scattering solutions 
	$$
	\delta u(x) \propto \hat{\cal{A}} \exp(i k x) = (-ik + \tanh(x)) \exp(i k x)
	$$
with $\omega=\pm\sqrt{k^2+1}$ are also well known.


\section{Linearization of\\Korteweg-de Vries equation} It is tempting to suggest that the equations for small excitations around solitons of integrable PDEs are candidates for new, previously unknown instances of scattering without reflection. To that end, we have found that the Korteweg-de Vries equation
	$$
	u_t - 6uu_x + u_{xxx} = 0\,,
	$$
can be linearized around the stationary soliton
	$$
	\bar{u}(x) = -2\sech^2\!x + \nicefrac{2}{3}
	$$
using the approximate solution
	$$
	u(x,\,t) \approx \bar{u}(x) + \delta u(x,\,t)
	$$
to generate $\delta u_t = \hat{H}\,\delta u$, where
	$$
	\hat{H} = -\px^3 + 6\bar{u} \px + 6 \bar{u}_x\,.
	$$
As expected, the Liouvillian $\hat{H}$ is reflectionless, which can be verified by intertwining it, as in equation (\ref{intertwining_relationship}) again, with a scatterer-free Liouvillian:
	\begin{align*}
	\hat{H}_0 &= -\px^3+4\px \,,
	\end{align*}
using the intertwiner
	\begin{align*}
	\hat{\cal{A}} &= \px^4-4\tanh x\,\px^3 + \left(4-8\sech^2\!x\right)\px^2 \\
		&\hphantom{={}}\qquad\qquad\qquad\qquad + 8\sech^2\!x\,\tanh x\,\px \,.
	\end{align*}
The scattering solutions can be obtained from the intertwiner above; they are $\delta u(x,\,t) = \delta u(x) \exp(i \omega t)$, with
	\begin{align*}
	&
	\delta u(x) \propto \hat{\cal{A}} 
	\exp(i k x) = k\left[ 8 \sech^2(x) (k+i \tanh(x)) 
	\right.
	\\
	&\qquad\qquad
	\left.
	+k (k^2+4 i k \tanh(x)-4)\right] \exp(i k x) \,,
	\end{align*}
and $\omega = k^3 + 4 k$.


\section{Linearization of\\Liouville's equation} The integrability of the NLS, sine-Gordon, and KdV equations studied above all originate from inverse scattering, and thus fall under the paradigm of S-integrability \cite{solitons_book}. Let us test our conjecture now on Liouville's equation, which is instead C-integrable---by a change of variables. Liouville's equation is
	\begin{align}
	u_{xx} - u_{tt} = e^{-u}\,,\label{Liouville_equation}
	\end{align}
and it can be reduced to a linear PDE through an appropriate change of variables \cite{PDE_book__polyanin_and_zaitsev}. Its most general solution has also been obtained in \cite{crowdy1997_141}, using elementary methods. We take the stationary soliton
	\begin{align}
	\bar{u}(x)= -\ln(2 \sech^2(x))
	\label{Liouville_soliton}
	\,,
	\end{align}
that belongs to a broad class of solutions \cite{bhutani1992_1049},
	\begin{align}
	&u(x,\,t)= -\ln\left(-8 f'(x+t)g'(x-t) \right. 
	\label{bhutani_solution}
	\\	
	&\qquad\qquad\qquad \left. \times \sech^2(f(x+t)+g(x-t)+d)\right)
	\,,
	\nonumber
	\end{align}
where $f(\xi_+)$ and $g(\xi_-)$ are arbitrary functions of one variable, and $d$ is an arbitrary constant \cite{Liouville_csch}. The stationary soliton in (\ref{Liouville_soliton}) can be recovered by choosing $f(\xi_+)=\xi_+/2$, $g(\xi_-)=-\xi_-/2$, and $d=0$.

Let us perturb the soliton in (\ref{Liouville_soliton}) as
	$$
	u(x,\,t) = \bar{u}(x) + \delta u(x)\,e^{i\omega t}.
	$$
Performing this linearization leads to
	$$
	\bigl( -\px^2 - 2\sech^2\!x \bigr)\, \delta u  = E \, \delta u ,
	$$
with $E=\omega^2$ and $\omega=k^2$, which is the P\"oschl-Teller equation (\ref{PT}) with $n=1$, so there will again be no reflection of the scattering waves.

Physically, the reflectionless property of the linearized Liouville's equation is necessary to ensure that small localized packets can penetrate the main soliton without reflection. For example, a rightward moving packet will be given by expression (\ref{bhutani_solution}) with $f(\xi_+)=\xi_+/2$, $g(\xi_-)=-\xi_-/2 + \delta g(\xi_- + t_{0})$, and $d=0$, where  $\delta g(\xi_-)$ is a small packet localized at $\xi_- =0$ and $t_{0}$ is the moment of collision between the packet and the main soliton.

The four nonlinear PDEs that have been treated so far---Bogoliubov-de Gennes, sine-Gordon, Korteweg-de Vries, and Liouville's equation---are all integrable systems that confirm our conjecture that {\it the linearization of an integrable PDE around a soliton should produce scattering without reflection at all energies.} Although this conjecture is grounded firmly in the ability of solitons in integrable systems to penetrate each other without loss of identity, it is natural to ask if it could be applied to stationary solutions of a non-integrable PDE. To provide this contrasting study, we will now examine the behavior of two non-integrable systems under the same linearization scheme that has been utilized so far.


\section{Linearization of a\\sawtooth-Gordon equation} The first non-integrable system to be examined is a sawtooth-Gordon system, defined by
	\begin{align*}
	&u_{xx} - u_{tt} = F(u),\\
	&F(u) = \tfrac{1}{4}u - \left\lfloor \tfrac{1}{4}(u + 2) \right\rfloor ,
	\end{align*}
where $\left\lfloor\ldots\right\rfloor$ denotes the floor function. This equation admits a stationary solution of 
	$$
	\bar{u}(x) =
		\left\{
  		\begin{array}{ll}
    	2 e^{x/2} &\text{for } x < 0\\
    	4 - 2 e^{-x/2} &\text{for } x \geq 0
  		\end{array}
  	\right. .
	$$
Moving kink and anti-kink solutions can be generated from the stationary solution above through Lorentz transformations. Numerical simulations of collisions between a moving kink and an anti-kink solution result in the annihilation of both entities and the production of radiative ripples that confirm this sawtooth-Gordon equation is indeed non-integrable. Using a linearization of
	$$
	u(x,\,t) \approx \bar{u}(x) + \delta u(x)\,e^{i\omega t},
	$$
we find that the perturbation will obey
	$$
	\bigl( -\px^2 - \delta(x) \bigr)\, \delta u  = E\,\delta u ,
	$$
with $E=\omega^2 - \tfrac{1}{4}$, which is just a generic scattering problem with a $\delta$-function potential. It is well known to produce reflections at all finite energies.


\section{Linearization of a $\phi^4$ model} The second non-integrable system we will examine is a $\phi^4$ model with the equation of motion
	$$
	\phi_{xx} - \phi_{tt} = 2\phi^3 - 2\phi,
	$$
that admits the stationary solution
	$$
	\bar{\phi}(x) = \tanh(x).
	$$
Lorentz transformations of the stationary solution above can produce moving kink and anti-kink solitary wave solutions. The behavior of kink and anti-kink collisions in the $\phi^4$ model is rich with detail \cite{campbell1983_1}. For the current study however, we only confirm the non-integrability of the system by numerically simulating a kink and anti-kink collision. With the kink and anti-kink each moving at half the speed of light, their collision results in the reflection of both entities and outward radiating ripples. Continuing now with a linearization of
	$$
	\phi(x,\,t) \approx \bar{\phi}(x) + \delta \phi(x)\,e^{i\omega t},
	$$
we find that the perturbation will obey
	$$
	\bigl( -\px^2 - 6\sech^2 x \bigr)\, \delta\phi  = E\,\delta\phi ,
	$$
with $E=\omega^2 - 4$. This is a P\"oschl-Teller potential with $n=2$, and is thus reflectionless.

From the analysis of the non-integrable sawtooth-Gordon equation and $\phi^4$ model above, we can see that the result of linearizing around stationary solutions of a non-integrable system is quite different from that of linearizing around a stationary solitonic solutions of an integrable system. In particular, linearized non-integrable systems may be reflective or reflectionless, while linearized integrable systems are consistently reflectionless.


\section{Summary and outlook}

Based on a close study of the \BdG equation---a linearization of an attractive nonlinear Schr\"{o}dinger equation---we conjecture that the linearization of physically relevant integrable PDEs around stationary solitons should produce linear problems that demonstrate scattering without reflection at all energies. This prediction is verified in the sine-Gordon, Korteweg-de Vries, and Liouville's equations. The last case indicates that this phenomenon spans two paradigms of integrability: S-integrability for the nonlinear Schr\"{o}dinger, sine-Gordon, and Kortweg-de Vries equations, and C-integrability for Liouville's equation. The applicability of our conjecture towards both S-integrable and C-integrable systems suggests that the emergence of reflectionless scattering from the linearization of integrable PDEs is deeply connected to integrability {\it per se}.

We indicate that the transparency of linearized PDEs may be necessary to ensure that they correctly predict the transparency of large solitons to small solitons, small breathers, or small packets (for Liouville's equation) at the level of the original integrable nonlinear PDEs. It remains to be verified that the observed transparency persists for multi-soliton solutions and for other integrable PDEs in general.

For a contrasting study, we also linearized stationary solutions of non-integrable systems to produce both reflective and reflectionless equations, suggesting that there exists factors other than integrability that can produce reflectionless equations. The nature of this connection remains to be investigated however.


\begin{acknowledgments}
We are grateful to Vanja Dunjko, Steven Jackson, Alfred No\"{e}l, Shabnam Beheshti, Wen-Xiu Ma, Dmitry Pelinovsky, and David Campbell for enlightening discussions on the subject. This work was supported by grants from the Office of Naval Research {\it (N00014-12-1-0400)} and the National Science Foundation {\it (PHY-1402249)}.
\end{acknowledgments}


\bibliography{Nonlinear_PDEs_and_SUSY_014,Bethe_ansatz_012,BdG_SUSY_paper}

\begin{thebibliography}{21}
\bibitem [{non()}]{non_uniquiness}%
  \BibitemOpen
  \href@noop {} {}\bibinfo {note} {The intertwiner (\ref{BdG_intertwiner}) is
  not unique; in particular, the intertwining property will be preserved when
  the intertwiner $\hat{\cal A}$ is multiplied from the left by any polynomial
  of $\hat{H}$ and from the right by any polynomial of $\hat{H}_{0}$. It
  appears, however, that the intertwiner (\ref{BdG_intertwiner}) is the
  simplest member of its family.}\BibitemShut {Stop}%
\bibitem [{com()}]{comment_on_antiparticles}%
  \BibitemOpen
  \href@noop {} {}\bibinfo {note} {The negative energy eigenstates of
  $\hat{H}_{0}$, $ w_{k,\,0,\,-}(x) = (0,\,1)^\intercal\exp(i k x) $, lead to
  an identical set of scattering solutions $\delta\psi(x,\,t)$ for the BdG
  equation (\ref{BdG}), just with the roles of $u$ and $v$ interchanged. We
  discard these solutions as redundant.}\BibitemShut {Stop}%
\end{thebibliography}


\end{document}